# Effect of annealing temperature on exchange stiffness of CoFeB thin films

Jaehun Cho, Jinyong Jung, Shin-Yong Cho, Chun-Yeol You[*]

*Department of Physics, Inha University, Incheon 402-751, South Korea*

We investigate the exchange stiffness constants of 28-nm-thick CoFeB film using Brillouin light scattering. Series of CoFeB films are prepared on the MgO(001) substrate with or without additional 5-nm thick MgO buffer layer, the effect of the annealing temperature on the exchange stiffness constants are studied. We found that the exchange stiffness constant of 400°C annealed sample with MgO buffer increased by 10 % form the 200°C annealed sample (=0.73 ± 0.01 × $10^{-11}$ J/m), while the exchange stiffness constant of without MgO buffer layer sample increase by 6 % from the as-grown sample (=1.11 ± 0.02 × $10^{-11}$ J/m).

[*] Corresponding author. TEL.: +82 32 860 7667; FAX: +82 32 872 7562.

*E-mail address:* cyyou@inha.ac.kr

# I. INTRODUCTION

Magnetic tunnel junctions (MTJs) with a MgO barrier have been studied as a fundamental building block of non-volatile STT-MRAM (spin transfer torque magnetic random access memory) devices [1][2]. Moreover, the magnetic properties of CoFeB are being actively studied as candidate materials for the STT-MRAM. Since CoFeB/MgO/CoFeB multilayer shows huge tunneling magneto-resistance (TMR) ratio [3] and perpendicular magnetic anisotropy [4], the basic magnetic properties of the CoFeB/MgO/CoFeB multilayer are important in the design of STT-MRAM. For example, recently, we theoretically reported on that the switching current density of the STT-MRAM is very sensitive to on the exchange stiffness constant of the free layer [5]. However, only a few experimental measurements have been reported for the exchange stiffness constant of CoFeB [6][7][8], despite the exchange stiffness constant is a fundamental physical quantity and related with the number of nearest neighbor magnetic atoms and Curie temperature of materials [9], and it is important in the domain wall motion devices because it will determine the domain wall width and energy [10]. In the CoFeB/MgO/CoFeB multilayer stacks, it is well-known that careful annealing process is require in order to obtain huge TMR, because the formation of crystalline phase of CoFeB. Usually, CoFeB layer is sputtered as an amorphous phase on the bottom layer and crystallized by post annealing process, during the annealing process, CoFeB crystallizes at the MgO/CoFeB interface and B diffuses from the MgO/CoFeB interface to the Ta capping layer. Miyajima *et al.* [11] has reported that crystallization of CoFeB layers is strongly dependent on the capping material due to the B diffusion, where B mainly diffused to the capping layers and rarely to the MgO layers with increasing temperature. Because magnetic properties such as the saturation magnetization, perpendicular magnetic anisotropy (PMA), and the exchange stiffness constant are sensitive function of the non-magnetic B composition, they are affected by the annealing processes.

In our previous work, the exchange stiffness constant of CoFeB film on Si substrate decreases from 1.41 to 0.98 × $10^{-11}$ J/m with the fabrication conditions such as Ar gas pressure.[12] The stiffness constant is proportional to the product of the number of nearest neighbor magnetic atoms.[13] We claim that the atomic concentrations of Fe and Co are changed with the Ar gas pressure. We confirm this result using X-ray photoelectron spectroscopy (XPS) as the different atomic concentration of Fe and Co. In this study, we investigate the dependence of the magnetic properties of $Co_{40}Fe_{40}B_{20}$ fim on MgO(100) substrate such as the saturation magnetization and the exchange stiffness constants on annealing temperature by using Brillouin light scattering (BLS). Furthermore we discuss the effect of the additional MgO buffer layer between MgO(001) substrate and the $Co_{40}Fe_{40}B_{20}$ ferromagnetic layer. By the analysis of the surface and bulk spin wave (SW) modes, we determined the saturation magnetization and exchange stiffness constants for series of $Co_{40}Fe_{40}B_{20}$ samples. We found that the saturation magnetization and the exchange stiffness constant are sensitive on the fabrication conditions, with or without MgO buffer layer. The observed results can be explained by the different B concentration for with or without MgO buffer layer samples using the XPS depth profile measurements.

## II. EXPERIMENTS

Series of 28-nm $Co_{40}Fe_{40}B_{20}$ samples were prepared on MgO(001) substrate using a dc magnetron sputtering system. Hereafter, the films grown with 5-nm-thick additional MgO buffer layer and without MgO buffer layer on the MgO(001) substrate are labeled w/ MgO and w/o MgO buffer layer, respectively. Schematic structures of w/ and w/o MgO buffer layer samples are shown in the inset of Fig. 1 (a) and (b). We used $Co_{40}Fe_{40}B_{20}$ alloy target with purity of 99.99 % and the deposition was carried out at 100 W under a base pressure of $3 \times 10^{-8}$ Torr or lower. The deposition temperature of the substrate was room temperature. More details of sample fabrication conditions are found elsewhere [12]. The thickness of $Co_{40}Fe_{40}B_{20}$ thin films was kept at 28 nm with 4-nm-thick Ta capping layer.

Role of the Ta capping layer will be discussed later. After films growth, we annealed the sample at 200, 300, and 400 ºC for 1 hour which is used different pieces of the sample as the temperature increasing ratio was 5 ºC/s, which is much faster than typical annealing condition (0.25 ºC/s).

SW spectra were studied by a BLS system with Sandercock type (3+3) tandem Fabry-Perot interferometer [14,15]. The probe light source used a single longitudinal mode of 532 nm of diode-pumped solid state laser with an output power of 300 mW. Backscattering geometry was used to observe the light scattered by thermal excitations with an in-plane SW wavevector $q_{//} = 1.670 \times 10^5$ cm$^{-1}$ with the incident angle of 45º. External magnetic fields of up to $6 \times 10^3$ $k$A/m were applied parallel to the film plane and perpendicular to the scattering plane [16]. The accumulation time for each spectrum was about 60 min., all measurements were performed *ex*-situ at room temperature.

**III. RESULT and DISCUSSION**

The magnetic hysteresis loops measured with the applied magnetic field parallel to the film plane are presented in Fig. 1(a) for the sample w/o MgO buffer layer and Fig. 1(b) for the sample w/ MgO buffer layer (see the insets). It is clearly shown that w/o MgO buffer layer samples have smaller coercivity (~ 1.5 kA/m) and sharp switching for whole series of samples, while w/ MgO buffer layer samples have little larger coercivity (~ 2 kA/m). Since all samples are nominally composed of the same materials, such difference switching mechanisms indicate that there are noticeable changes in the microstructures in CoFeB layer because of the MgO buffer layer. Even though we used the same MgO as buffer layer on the top of the MgO(001) substrate, the microstructure of the MgO buffer layer are quite different from the single crystalline substrate, therefore the difference between w/ and w/o MgO buffer layer is understandable.

Fig. 2 shows typical SW spectra obtained from 28-nm CoFeB film as-deposited w/o MgO buffer layer. Two spectra for the applied filed of $0.30 \times 10^2$ and $1.91 \times 10^2$ kA/m, are plotted together in order to show the magnetic field dependences of the excited SW frequencies. The stronger peaks labeled DE observed on the positive frequency shift region (anti-Stokes) have their origin in scattering from the surface wave. For the ferromagnetic thin film, the surface wave is known as the Damon-Eshbach (DE) mode [17]. The peaks labeled $B_1$ are the first-order bulk SW modes [18]. Due to analyze the separate peak position between the DE mode and the first bulk mode, 28-nm-thick sample is taken. In order to determine the magnetic properties, the SW frequencies are measured as a function of the applied magnetic field. The results are displayed in Fig. 3. The dependences of the DE and the first bulk modes on the applied magnetic field have been used to obtain the exchange stiffness constants and the saturation magnetization with the excited SW frequencies with following equations [19]:

$$f_{DE} = \frac{\gamma}{2\pi}\left[H\left(H + 4\pi M_s\right) + \left(2\pi M_s\right)^2 \left(1 - e^{-2q_{//}d}\right)\right]^{1/2}, \qquad (1)$$

$$f_{Bulk} = \frac{\gamma}{2\pi}\left[\left(H + \frac{2A_{ex}}{M_s}\left(q_{//}^2 + \left(\frac{n\pi}{d}\right)^2\right)\right) \times \left(H + \frac{2A_{ex}}{M_s}\left(q_{//}^2 + \left(\frac{n\pi}{d}\right)^2\right) + 4\pi M_s\right)\right]^{1/2}. \qquad (2)$$

Where $H$ is the applied magnetic field, $\gamma$ is the gyromagnetic ratio ($\gamma = g|e|/2mc$, where $g$ is the spectroscopic splitting factor, $e$ is the charge of electron, $m$ is the mass of electron and $c$ is the velocity of the light.), $d$ is the ferromagnetic layer thickness, $n$ is the order number for the bulk modes, $M_s$ is the saturation magnetization, and $A_{ex}$ is the exchange stiffness constant. Here, we ignore the crystalline anisotropy energy, because of amorphous nature of CoFeB layer. Furthermore, the surface anisotropy is also ignored because the CoFeB thickness is ~28 nm. In this thickness range, the surface energy term is negligible due to the inverse proportionality of the surface energy term [20]. The gyromagnetic ratio and saturation magnetization can be determined by fitting with Eq. (1) and the exchange stiffness constant from Eq. (2). The experimentally obtained SW spectra are represented by black open squares (DE mode) and red open circles (the 1st bulk mode) and the black and red solid lines are the best fits of the experimental data with Eq. (1) and (2). A good agreement between fitting results and the experiments are found for DE and bulk modes. From the fitting procedures, we determined $M_s$ and $A_{ex}$ while $g$ value is fixed as 2.13.

The saturation magnetizations obtained from the BLS measurements are plotted in Fig. 4 as a function of the annealing temperature. The black open squares are w/o MgO buffer layer samples and the red open circles are w/ MgO buffer layer samples. As shown Fig. 3, the saturation magnetization has no tendency in the w/o MgO samples however, the saturation magnetization has a minimum value of about 820 kA/m at annealing temperature is 200 ºC and increases with increasing the annealing temperature. Those values of the saturation magnetization of the as-deposited w/o MgO and w/ MgO buffer layer samples are 1000 kA/m and 890 kA/m, respectively. We speculate that the annealing processes of w/o MgO buffer layer samples are not changed amorphous CoFeB layer microstructure significantly, despite of the annealing. By the way the annealing processes of w/ MgO buffer layer promote microstructural transformation (i.e. crystalline) of amorphous CoFeB layers [21]. The noticeable difference of the saturation magnetization between w/o MgO samples and w/ MgO samples will be discussed later.

The determined exchange stiffness constants by BLS measurements as a function of annealing temperature are plotted in Fig. 5. We find that the exchange stiffness constants varied from $1.11 \pm 0.02$ to $1.18 \pm 0.01 \times 10^{-11}$ J/m and $0.73 \pm 0.01$ to $0.81 \pm 0.02 \times 10^{-11}$ J/m by annealing temperature for w/o MgO buffer layer samples and w/ MgO buffer layer samples, respectively. The values of the exchange stiffness constants of w/o MgO buffer layer samples are similar except annealed temperature 400 ºC. However, the exchange stiffness constant of 400 ºC annealed sample increased by 6 % from as-grown sample. The other hands the values of the exchange stiffness constants of w/ MgO buffer layer samples increase with increasing the annealing temperature. The difference of maximum and minimum exchange stiffness constants in w/MgO buffer layer sample is about 10%. This is little bit larger than the difference of the w/o MgO buffer layer samples. We already have reported that the small change of the exchange stiffness constants may cause the big change of the critical switching current density because the exchange length of CoFeB is comparable with the single MTJ size (~30 nm). Especially, the abrupt changes are expected around $1.0 \times 10^{-11}$ J/m, 10 % variation of the exchange constant may indicate (quantitatively) what is the significant changes in switching current. (see Fig. 1 of Ref. [5]).

In order to check the atomic concentration of each material (Co, Fe and B), we measure x-ray photoelectron spectroscopy (XPS) depth profile of w/o MgO and w/MgO buffer layer samples for each annealing temperature. Fig. 6(a) and (b) show the changes of relative Fe, Co, B, and Ta atomic compositional ratios of the w/o and w/ MgO buffer layer CoFeB samples as a function of etched time, respectively. The approximate positions of each layer regions are indicated by two blue vertical lines.

H. Bouchikhaoui *et al.* [22] reported that the different concentration of B atoms at the CoFeB and Ta interface after isothermal and isochronal annealing mechanism in the varying annealing temperatures. In the as-deposited state, CoFeB layer and Ta capping layer appear well-separated by a sharp interface. After annealing process, B concentration has changed at the CoFeB/Ta interface. The amount of B located at the interface increases with annealing temperature. They asserted that the B concentration is increase at the CoFeB/Ta interface with the annealing temperature. Continuously now we show the w/o MgO buffer layer (Fig. 6(a)) cases, the atomic composition of B is high concentration and it is getting smaller as the increasing etching time in the as-grown sample. The depth profiles of B composition show similar tendency for all annealing temperatures. The ferromagnetic materials (Co, Fe) have almost the same composition regardless the annealing temperature for whole samples as shown in Fig. 6 (a) and (b). Therefore, we can claim the B diffusion is not changed with the annealing temperature due to the very rapid thermal annealing process (5 ºC/s).This result is similar to w/ MgO buffer layer (Fig. 6 (b)). It must be noted that our annealing processes are different from the isothermal and isochronal annealing, our results are different from Bouchikhaoui's [22]. In comparison with the CoFeB regions of w/ and w/o MgO buffer layer samples, the atomic composition of Co and Fe in w/o MgO buffer layer samples have larger atomic concentration of Co and Fe than w/ MgO buffer layer samples. Furthermore, the atomic composition of B in w/o MgO buffer layer samples is getting smaller as the increasing etching time. This B concentration gradient may cause the changes of the magnetic properties such as the saturation magnetization and exchange stiffness constant. Because when the non-magnetic B atoms are placed between magnetic atoms, it cuts the exchange coupling between two magnetic atoms. As a result, the number of nearest neighborhood is reduced, so that the exchange stiffness constant decreases [13].

## IV. SUMMARY

We have characterized the magnetic properties of MgO/CoFeB with/without the additional MgO buffer layers by varying the annealing temperatures. We found the exchange stiffness constant of 400 ºC annealed sample without MgO buffer increased by 6 % from room temperature sample while exchange constant of 400ºC annealed sample with MgO slightly increased by 10 % from 200 ºC annealed sample. Also, we investigate the changing of the exchange stiffness constant caused by the additional MgO buffer layer.

## ACKNOWLEDGEMENT

This work was supported by by INHA UNIVERSITY Research Grant.

## REREFENCES

Fig. 1. Typical magnetic hysteresis loops for 28-nm-thick CoFeB film with difference annealing temperature of (a) without MgO buffer layer (b) with MgO buffer layer, resistively. The inset shows the schematic view of without MgO buffer layer (a) and with MgO buffer layer (a), resistively.

Fig. 2. Typical BLS spectra record from 28-nm-thick CoFeB film as-deposited w/o MgO buffer layer when the angle of incident was 45º. The peak labeled "DE" is the Damon-Eshbach mode and "$B_1$" is the first bulk mode.

Fig. 3. Variation of SW frequency with an applied field for CoFeB sample with as-deposited w/o MgO buffer layer. The open black squares are Damon-Eshbach mode and the open red circles are 1st order bulk mode. The lines are fitted curves.

Fig. 4. Saturated magnetization of CoFeB films as a function of annealing temperature. The open black squares are w/o MgO buffer layer and the open red circles are w/ MgO buffer layer.

Fig. 5. Exchange stiffness constant of CoFeB films as a function of annealing temperature. The open black squares are w/o MgO buffer layer and the open red circles are w/ MgO buffer layer.

Fig. 6. XPS depth profiles of each materials (a) w/o MgO buffer layer and (b) w/ MgO buffer layer for 28-nm CoFeB films with the annealing temperatures. The approximate positions of the each layer regions are indicated by blue vertical lines.

Fig. 1.

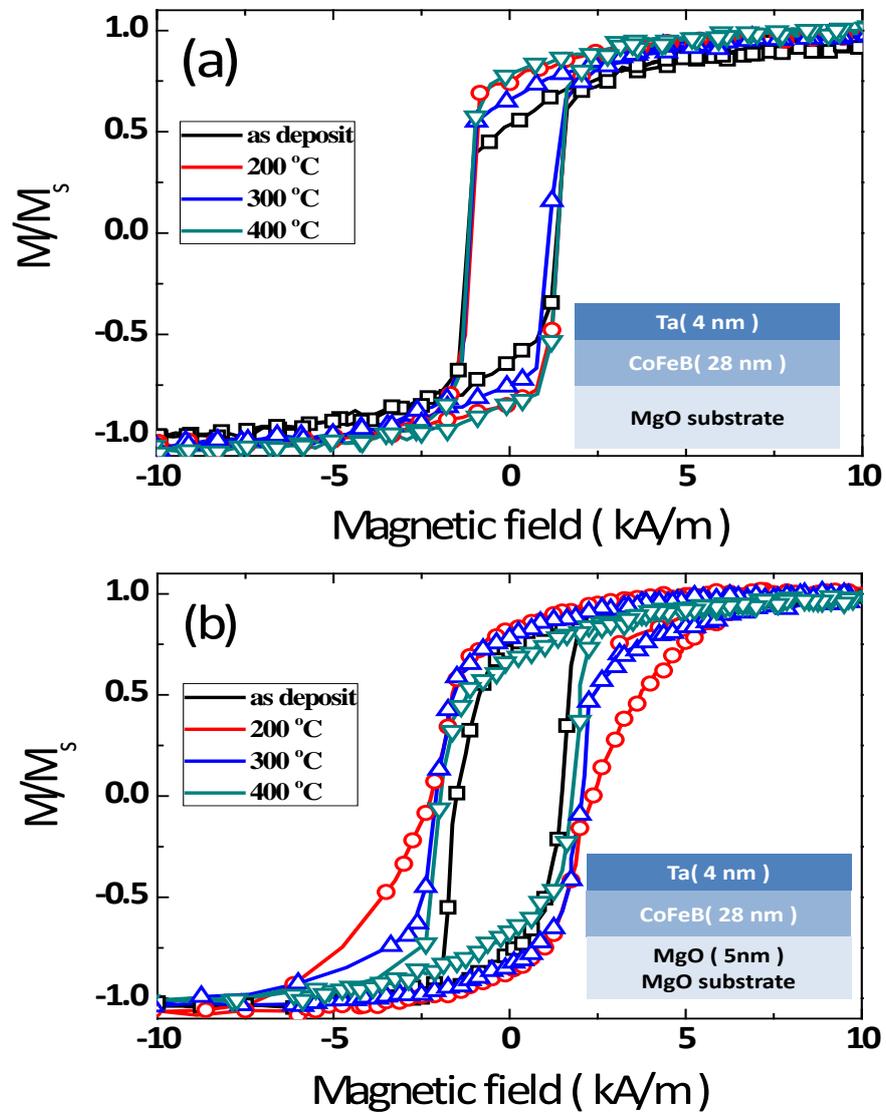

Fig. 2.

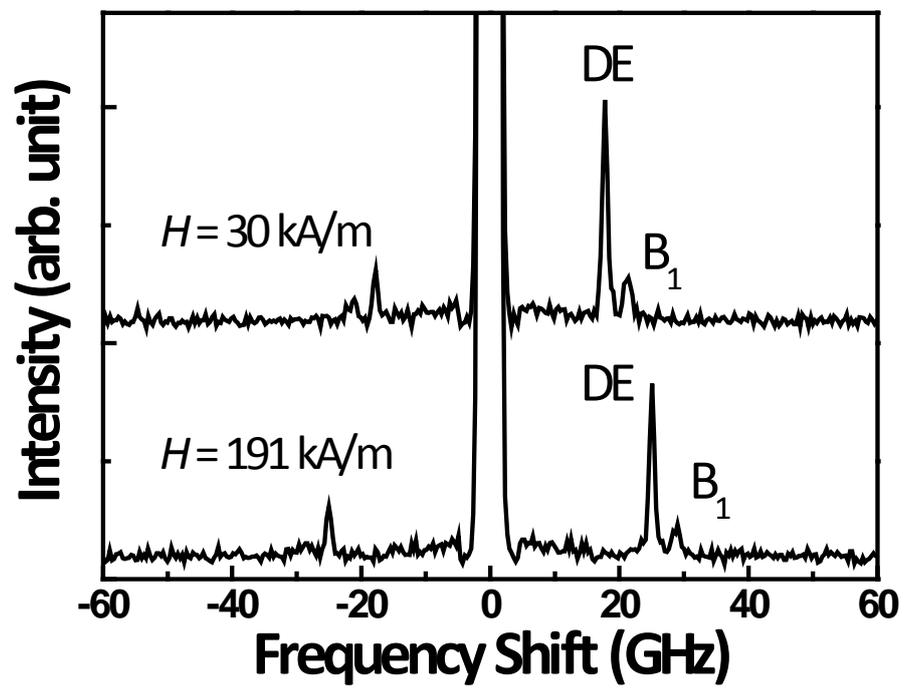

Fig. 3.

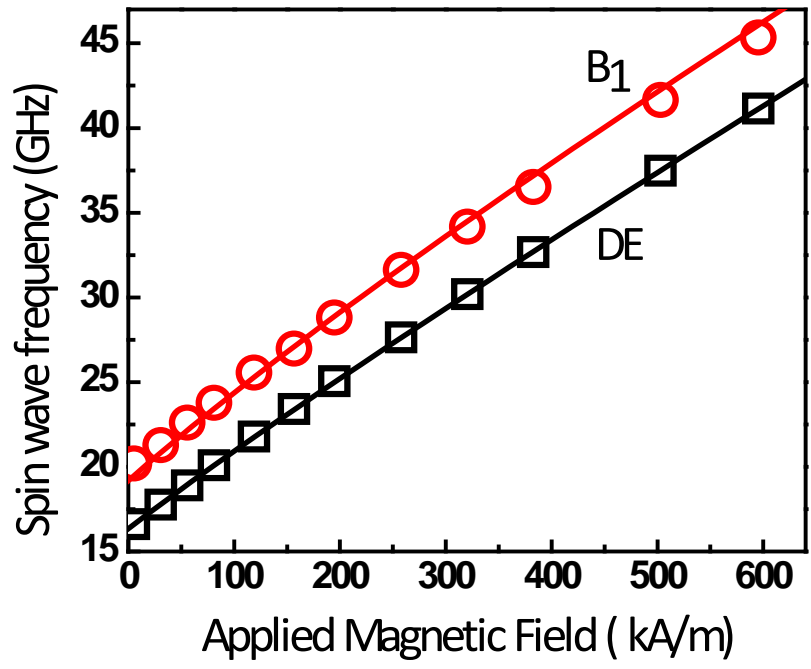

Fig. 4.

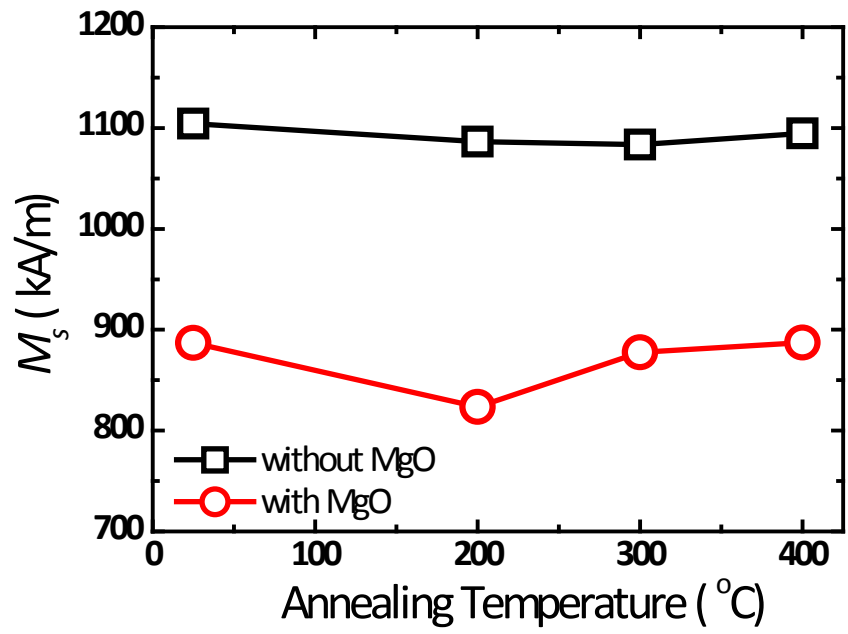

Fig. 5.

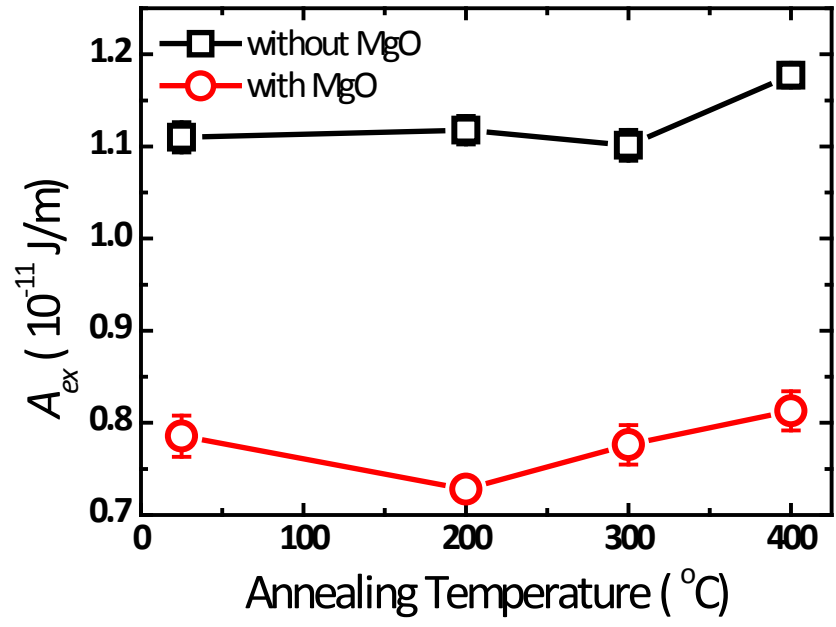

Fig. 6.

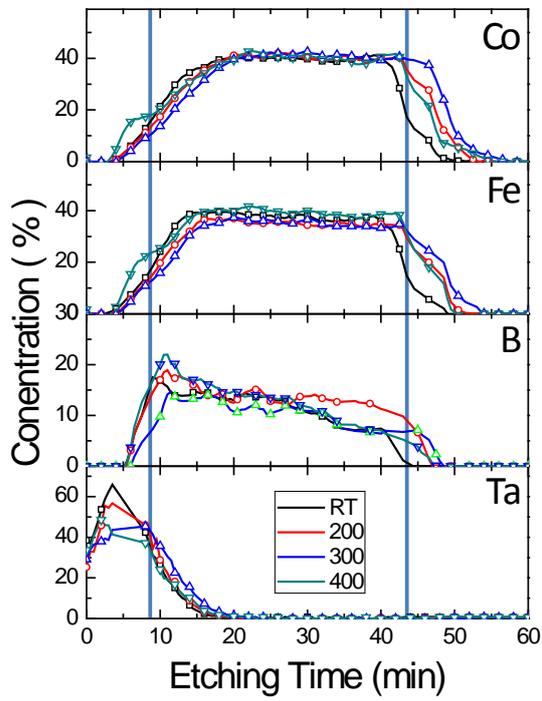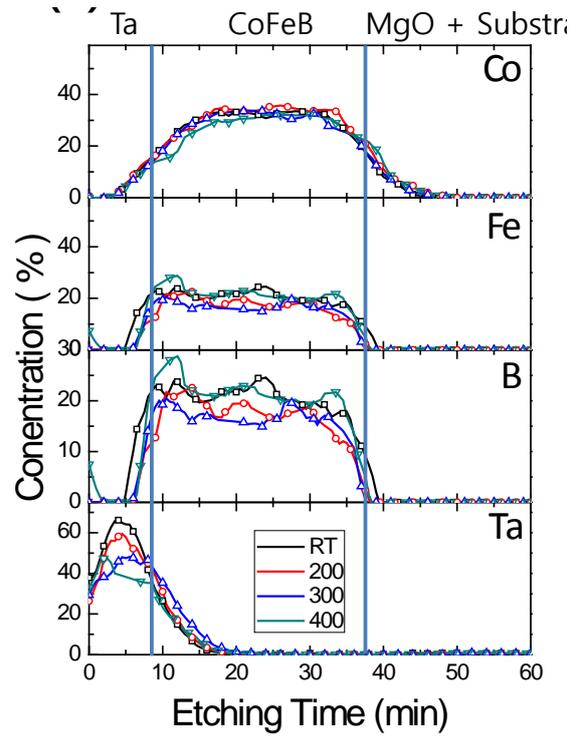